\input{aipcheck}

\documentclass[
    ,final            
  ]
  {aipproc}

\layoutstyle{6x9}

\begin{document}

\title{Four-Quark Mesons}


\author{L. Maiani}{
  address={Universit\`{a} di Roma `La Sapienza' and I.N.F.N., Roma, Italy}
}

\author{F. Piccinini}{
  address={I.N.F.N. Sezione di Pavia and Dipartimento di Fisica Nucleare 
e Teorica, via A.~Bassi, 6, I-27100, Pavia, Italy}
}

\author{A.D. Polosa}{
  address={Dipartimento di Fisica and I.N.F.N., Bari, Italy}
}
\author{V. Riquer}{
    address={CERN Theory Department, CH-1211, Switzerland
  }
}

\begin{abstract}
The features of a model interpreting the light scalar mesons 
as diquark-antidiquark bound states and the 
consequences of its natural extension to include 
heavy quarks are briefly reviewed. 
\newline\newline
{\bf Preprint numbers:} FNT/T-2004/22, BA-TH/503/04, CERN-PH-TH/2004-241.
\end{abstract}

\maketitle


The $q\bar{q}$ assignment has never really worked for the 
scalar mesons below 1~GeV. 
Alternative identifications have 
been proposed in the past~\cite{close}, notably the $f$ as a bound 
$K\bar K$ molecule~\cite{isgur} or as a $(q)^2(\bar q)^2$ state~\cite{jaffe}.
We illustrate in this contribution the 
hypothesis, examined in Ref.~\cite{mppr1}, that the lowest lying scalar 
mesons are $S-$wave bound states of a diquark-antidiquark pair. 
Following Ref.~\cite{jaffe-wilczek}, 
the diquark is more likely bound in the ${\bf \bar{3}_c}$, ${\bf 0_s}$
(color antitriplet, spin zero) channel. If strange 
quarks are included, Fermi statistics favors the ${\bf \bar{3}_f}$ combination.
Therefore $(q)^2(\bar q)^2$  states form a flavor $SU(3)$ nonet. 
We propose to put the $\sigma(450)$~\cite{kloe} in the 
$I=S=0$ state, 
and to assign to the $S=\pm 1$ states the $\kappa(800)$,  a $K\pi$ 
resonance seen by several experiments, 
most recently in the $K \pi \pi$ spectrum from $D$ decays~\cite{e791}. 
A simple hypothesis on the way the $(q)^2(\bar q)^2$ states may transform 
into a pair of pseudo-scalar mesons is found to give a rather good 
one-parameter description of the decays allowed by 
the OZI rule~\cite{ozi}. 
The extension of the picture to states including one ore more 
heavy quarks gives quite interesting predictions, accommodating 
recently discovered narrow states. 

\paragraph{Quantum numbers and spectrum}

We denote by $[q_1q_2]$ the fully antisymmetric state of the two quarks 
$q_1$ and $q_2$. 
The composition of the members of the nonet is as follows:    
\begin{eqnarray}
a^+(I=1,I_3=+1, S=0) &=& [s u] [\bar s \bar d] \nonumber \\
a^0(I=1,I_3=0, S=0) &=& \frac{1}{\sqrt{2}}\left( [s u] [\bar s \bar u] - 
[s d][\bar s \bar d] \right) \nonumber   \\
a^-(I=1,I_3=-1, S=0) &=& [s d] [\bar s \bar u] \nonumber \\
f_\circ(I=0, S=0) &=& \frac{1}{\sqrt{2}} \left( [s u] [\bar s \bar u] 
+ [s d] [\bar s \bar d] \right)  \nonumber \\
\sigma_\circ(I=0, S=0)&=& [u d] [\bar u \bar d] \nonumber  \\
\kappa (I=1/2, I_3=+1/2, S=+1) &=& [u d] [\bar s \bar d] \nonumber \\
\kappa (I=1/2, I_3=-1/2, S=+1) &=& [u d] [\bar s \bar u] \nonumber \\
\kappa (I=1/2, I_3=+1/2, S=-1) &=& [u s] [\bar d \bar u] \nonumber \\
\kappa (I=1/2, I_3=-1/2, S=-1) &=& [d s] [\bar d \bar u] \nonumber
\end{eqnarray}
where the neutral states $f(980)$ and $\sigma(450)$ are superpositions 
of the isoscalar states $f_\circ$ and $\sigma_\circ$. The mixing angle 
results to be small because the OZI rule is respected in the
physical mass spectrum.
 
In the limit of exact octet symmetry, the states given above are 
mass eigenstates, the mass matrix parameterized by $\alpha$ 
and $\beta$, the diquark masses squared with strange and non-strange 
content. In the most general case of octet symmetry breaking, two more 
parameters are required to account for symmetry breaking terms~\cite{mppr1}.  

The mass spectrum obtained is inverted with respect to what one would 
get for a  $q\bar q$ nonet: the isolated $I=0$ 
state is the lightest one and strange 
particles come next. The same pattern is shown by data and this 
is a most evident indication in favor of the 
four-quark nature of the scalar nonet.	

\paragraph{Strong decays}
Diquarks, being colored objects, cannot be separated 
by their anti-particles. As soon as the distance between 
two diquarks in a four-quark state
gets large enough, a $q-\bar q$ pair is created out of the vacuum and the 
state should dissociate into a baryon-antibaryon pair. 
This process is obviously kinematically forbidden as long as four-quark
light scalars are considered.

An alternative decay mechanism is the switching of 
a quark-antiquark pair between the two diquarks to form a 
pair of color neutral $q\bar q$ states (pseudoscalar mesons), 
which can indefinitely separate 
from each other. In the exact $SU(3)$ limit there is only one amplitude, 
${\cal A}$, to describe this process. The amplitude 
${\cal A}$ describes the tunneling from the bound 
diquark pair configuration to the meson-meson pair, made by the unbound 
final state particles. As seen in Ref.~\cite{mppr1}, the value 
${\cal A}= 2.6$~GeV gives a good description of the rates, compared 
to the available experimental information. 
The large value of ${\cal A}$ seems indicative of a short 
distance effect, making perhaps more justifiable the use 
of flavor $SU(3)$ symmetry. 
 
Our picture has some connection with baryonium 
states~\cite{rossi-veneziano} and with the 
$K\bar K$ molecule picture~\cite{isgur}. 
In the latter case, however, the analogy is only superficial. 

Adding the other three $SU(3)$ allowed (annihilation-)couplings 
(neglecting a fourth 
coupling related to a pure singlet-to-singlets amplitude) improves the 
description of the OZI allowed channels, except for the 
$\kappa$ width, which seems to be sensibly smaller than the observed one.
Also the OZI forbidden decay $f\to \pi \pi$, turns 
out to be too small with respect to the experimental rate, 
even allowing for the full $SU(3)$ effective strong decay
Lagrangian. 
It is quite possible that this 
mode proceeds via a different mechanism.

However the overall picture is encouraging and
reinforces considerably the case of the 
scalar mesons as $(q)^2(\bar q)^2$ states. 

\paragraph{Open and hidden charm mesons} 
A natural extension of the 
present scheme is the existence of analogous states where one or more 
quarks are replaced by charm or beauty. We consider the case of charm, 
extension to beauty is obvious. Open charm scalar mesons of the form
$S=[cq][\bar{q}\bar{q}]$, 
fall into characteristic ${\bf 6}\oplus \bar{{\bf 3}}$ multiplets of
$SU(3)_f$. The $\bar{\bf 3}$ has the same conserved quantum numbers
of $c\bar q$ states (`cryptoexotic'), 
but the ${\bf 6}$ has a pure exotic content. 
Hidden charm states of the form $[cq][\bar c \bar q]$ fall into
${\bf 8}\oplus {\bf 1}$ multiplets of $SU(3)$.
In Ref.~\cite{mppr1} a list of possible 
decays and related thresholds has been given. 

Two issues are crucial to the description of open or hidden
charm four-quark mesons: {\it isospin breaking} and 
{\it heavy-quark spin symmetry}. 
These aspects are briefly summarized 
in the next two paragraphs.

The mesons $a(980)$ and $f(980)$ are degenerate within about 
10~MeV~\cite{pdg}. This reflects the smallness 
of the OZI violating contributions (annihilation graphs)
to the mass matrix, which would align the mass eigenstates to pure $SU(3)$ 
representations. Thus sizeable deviations from the isospin basis 
are expected.
Due to asymptotic freedom, suppressing quark pair 
annihilation into gluons, we expect annihilation contributions
to be even smaller 
in systems containing heavy quarks. The mass eigenvalues 
will be aligned with states diagonal with respect to quark masses, 
even for the light, up and down, quarks~\cite{rossi-veneziano}.
The $D_{sJ}(2632)$~\cite{selex}, if confirmed, could be interpreted 
as a $[cd][\bar d \bar s]$ state, not an isospin eigenstate~\cite{mppr2},  
whose decay into $D^0K^+$ is OZI forbidden~\cite{ozi}.

The approximate spin-independence of heavy quark 
interactions, which is exact in the limit of infinite 
charm mass, implies both spin zero and spin one diquarks to form 
bound states. This implies a rich spectrum of states with $J=0,1,2$. 
The states with $J^{PC} = 1^{++}$ and $2^{++}$ could be 
identified~\cite{mppr3} 
with the $X(3872)$ and $X(3940)$ seen in BELLE data~\cite{BELLE}. 

Also for these states isospin breaking would apply. An indication of the 
latter phenomenon comes from the observation of the relative decay rate of
$X\to J/\psi + \rho$ and $X\to J/\psi + \omega$. 

Heavy-light diquarks can be the building blocks of a rich spectrum
of states which can accommodate some of the newly observed charmonium-like
resonances not fitting a pure $c\bar c$ assignment. A new charm spectroscopy 
could be behind the corner.

{\it FP wishes to thank G.M. Prosperi for his kind invitation. }

\end{document}